# JND-BASED PERCEPTUAL VIDEO CODING FOR 4:4:4 SCREEN CONTENT DATA IN HEVC

*Lee Prangnell and Victor Sanchez*

{l.j.prangnell, v.f.sanchez-silva}@warwick.ac.uk

Department of Computer Science, University of Warwick, England, UK

## ABSTRACT

The JCT-VC standardized Screen Content Coding (SCC) extension in the HEVC HM RExt + SCM reference codec offers an impressive coding efficiency performance when compared with HM RExt alone; however, it is not significantly perceptually optimized. For instance, it does not include advanced HVS-based perceptual coding methods, such as Just Noticeable Distortion (JND)-based spatiotemporal masking schemes. In this paper, we propose a novel JND-based perceptual video coding technique for HM RExt + SCM. The proposed method is designed to further improve the compression performance of HM RExt + SCM when applied to YCbCr 4:4:4 SC video data. In the proposed technique, luminance masking and chrominance masking are exploited to perceptually adjust the Quantization Step Size (QStep) at the Coding Block (CB) level. Compared with HM RExt 16.10 + SCM 8.0, the proposed method considerably reduces bitrates (Kbps), with a maximum reduction of 48.3%. In addition to this, the subjective evaluations reveal that SC-PAQ achieves visually lossless coding at very low bitrates.

## 1. INTRODUCTION

The transmission and storage of YCbCr 4:4:4 Screen Content (SC) video data is becoming increasingly ubiquitous in mission critical applications including medical imaging and video, telepathology and business-related video conferencing. For this reason, it is especially important to preserve the perceptual quality of the luma and chroma aspects of SC video data. In addition to this, it is highly desirable to minimize bitrates, as much as possible, in order to facilitate smooth data transmissions and also to reduce burdens related to bandwidth and data storage. Due to the significant importance of delivering high quality SC video data at low bitrates, JCT-VC has integrated advanced SCC coding tools for utilization in the HEVC HM RExt reference codec [1]; this standardized SCC extension is known as SCM [2]-[4].

The main SCC tools included in SCM comprise Intra Block Copy (IBC) [5], Palette Mode (PM) [6], Adaptive Motion Vector Resolution (AMVR) [7] and also Adaptive Color Space Transform (ACST) [8, 9]. Combined with the video coding algorithms already included in HM RExt, the aforementioned SCC coding tools achieve a high level of compression efficiency when applied to raw 4:4:4 SC video data. However, due to the absence of perceptual optimization in HM RExt + SCM, there is room for improvement in terms of facilitating further bitrate reductions without incurring a perceptually discernible decrease of visual quality in the reconstructed SC video data. The proposed technique provides a solution to this shortcoming.

As confirmed in document JCTVC-U1015 [10], the types of data provided by JCT-VC for testing SC techniques are as follows: Text and Graphics with Motion (TGM), Mixed Content (MC), Camera-Captured Content (CCC) and animations. For the purpose of research and testing, JCT-VC has provided YCbCr 4:4:4, YCbCr 4:2:0 and RGB versions of the same sequence [10]. It is well known that chroma subsampled 4:2:0 TGM data, for example, can cause significant issues concerning the readability of the textual content; in other words, the readability of text is much clearer in 4:4:4 SC video data [11]. This is potentially a major problem assuming that the SC data is being utilized in mission critical medical applications, for example. As such, we focus on YCbCr 4:4:4 data in this paper.

The IBC, PM, AMVR and ACST techniques in SCM are designed to maximally discard spatiotemporal and statistical redundancies in raw SC video data. These tools are especially useful for compressing SC data with repeating patterns and also in the mathematically lossless SCC modality. However, for lossy coding applications, SCM is still heavily reliant on the transform coding and the scalar quantization methods already present in HM.

The compulsory scalar quantization technique in HEVC HM is known as Uniform Reconstruction Quantization (URQ) [12]. URQ is not optimized for Human Visual System (HVS)-based perceptual quantization. URQ is designed to indiscriminately quantize transform coefficients in Y, Cb and Cr Transform Blocks (TBs) at equal levels according to the Quantization Parameter (QP). Note that Rate Distortion Optimized Quantization (RDOQ) [13, 14] and Selective RDOQ (SRDOQ) [15] are often optionally enabled in HM to work in conjunction with URQ. However, as confirmed in preliminary tests that we conducted, RDOQ and SRDOQ offer little to no improvement — in terms of bitrate reduction — when employed to perceptually quantize SC video data. Therefore, in this paper, we focus on the perceptual optimization of URQ for HVS-based quantization at the CB level.

Since 2013, several state-of-the-art JND-based perceptual video coding techniques have been proposed for the HEVC standard. These include the JND methods proposed by Naccari and Mrak [16], S. Wang et al. [17], Kim et al. [18], Bae et al. [19], Wu et al. [20] and Bae et al. [21]; note that all of these techniques can be employed with various types of video data including TGM, MC, CCC and animations. The JND methods proposed in [16]-[21] exploit HVS-based visual masking in the luma component of YCbCr video data, either by means of luminance-based perceptual quantization, transform coefficient-level perceptual optimization or a combination of the two. It is important to note, however, that these techniques do not account for chrominance JND, which constitutes a significant shortcoming. This is primarily due to the fact that chroma data typically contains considerable HVS-based psychovisual redundancies [22]; therefore, this should be exploited in JND-based and visually lossless coding techniques.

In this paper, we propose a novel CB-level JND-based perceptual quantization method for HEVC, which we name SC-PAQ. SC-PAQ refines Naccari's and Mrak's luma-only IDSQ technique in [16] whereby we account for the JND of chrominance data in addition to the JND of luminance data. Furthermore, the proposed technique accounts for the bit depth of the raw video data. SC-PAQ exploits both luminance masking and chrominance masking; luminance adaptation and chrominance adaptation are employed to achieve the desired perceptual masking. The primary objective of SC-PAQ is to attain a high compression performance without inducing a discernible decrease of visual quality in the reconstructed SC video data. SC-PAQ is particularly effective when applied to high bit depth YCbCr 4:4:4 SC video data including CCC data. The main reason for this is as follows: in high bit depth 4:4:4 sequences, data in the Cb and Cr channels typically consists of higher variances compared with the variances of data in the Cb and Cr channels in 8-bit chroma subsampled sequences.

The rest of this paper is organized as follows. Section 2 includes a detailed review of the relevant literature. Section 3 provides detailed technical expositions on the proposed SC-PAQ method. Section 4 includes the evaluation, results and discussion of the proposed technique. Finally, section 5 concludes the paper.

## 2. LITERATURE REVIEW

Mannos' and Sakrison's pioneering work in [23] formed a useful foundation for all frequency domain luminance CSF-based JND techniques which target HVS-based psychovisual redundancies in luminance image data. In [24], Ahuma and Peterson devise the first DCT-based JND technique, in which a luminance spatial CSF is incorporated. In [25], Watson expands on Ahuma's and Peterson's work by incorporating luminance masking and contrast masking into the DCT-based method; note that power functions corresponding to Weber's law are utilized in this method. Chou and Chen propose a pioneering pixel domain JND profile in [26], in which luminance masking and contrast masking functions are proposed for utilization in the spatial domain (luma data of 8-bit precision). This technique is based on average background luminance and also luminance adaptation. The authors further expand on this method in [27] by adding a temporal masking component, in which inter-frame luminance is exploited. X. Yang et al. in [28] propose a pixel-wise JND contribution to eradicate the overlapping effect between luminance masking and contrast masking effects. This technique also includes a filter for motion-compensated residuals. Zhang et al. in [29] propose a DCT based JND technique, in which a luminance adaptation parabolic piecewise function is derived. The authors established that perceptual compression based on JND modeling can be accomplished by employing luminance masking mechanisms for light and dark regions in luminance data.

In [30], Jia et al. propose a DCT-based JND technique founded upon a CSF-based temporal masking effect. Wei and Ngan in [31] introduce a DCT-based JND contribution for video coding, in which the authors incorporate luminance masking, contrast masking and temporal masking effects into the technique. Chen and Guillemot in [32] propose a spatial domain foveated masking JND technique, which is the first time that image fixation points are taken into account in JND modeling. More recently (i.e., from 2013 onwards), several JND based perceptual coding techniques have been proposed. In [16], Naccari and Mrak propose a JND-based and luma-only perceptual quantization scheme (IDSQ). It is designed to exploit luminance intensity masking; IDSQ is based on the research conducted by Zhang et al. in [29]. In [17], S. Wang et al. propose a spatiotemporal domain JND approach with application for SC data in HEVC. In this work, the authors employ Chou's and Chen's luminance masking and contrast masking functions previously proposed in [26, 27]. The overlapping effect proposed by X. Yang et al. in [28] is also utilized.

Kim et al. in [18] propose a hybrid frequency domain and spatiotemporal domain JND technique for HEVC. This method combines the following visual masking properties to create a JND visibility threshold: spatial CSF, luminance masking, luminance temporal masking and contrast masking. In [19], Bae et al. propose a perceptual video coding scheme for HEVC, in which the JND visibility threshold adapts to the size of the transform. As with previously proposed JND techniques, the JND profile in this method is based on luminance spatial CSF, luminance masking and contrast masking. Wu et al. in [20] propose a luminance-based JND model for images with pattern complexity. This method is based primarily on luminance contrast and pattern masking. In [21], Bae et al. propose a DCT-based JND technique which amalgamates luminance-based temporal masking and foveated masking. These visual masking effects are then combined with spatial CSF, luminance-based spatial masking and contrast masking to create a full JND profile. G. Wang et al. in [22] propose a multi-channel DCT-based JND technique for HEVC; this method combines spatial CSF, spatial masking, contrast masking and temporal masking.

In the overwhelming vast majority of previously proposed JND techniques, the perceptual relevance of chrominance data is typically neglected. This is an important shortcoming because, in general, chrominance data can be compressed to a significantly higher degree than luminance data. Furthermore, another issue with many previously proposed JND methods is the fact that they are designed for 8-bit precision only. In accordance with the literature review, there is presently a research gap in the literature for a JND technique which accounts for: i) Both the luminance channel and the chrominance channels; ii) The bit depth of raw video data — e.g., 8-bit and 10-bit YCbCr data; and iii) Evaluations on full chroma sampling video data (i.e., YCbCr 4:4:4 data). The proposed SC-PAQ technique satisfies these criteria.

## 3. PROPOSED SC-PAQ TECHNIQUE

In the proposed SC-PAQ technique, we extend Naccari's and Mrak's luminance masking-based perceptual quantization IDSQ technique in [16]. SC-PAQ improves upon IDSQ for the following reasons: chrominance masking is incorporated and, in addition, the bit depth of the raw luma data and the raw chroma data is taken into account. This equates to the fact that SC-PAQ is compatible with raw video data of any bit depth. Furthermore, luminance masking and chrominance masking parabolic piecewise functions are utilized. These JND-based functions adjust quantization levels; they facilitate modifications to the luma QStep and also the chroma Cb and Cr QSteps at the CB level.

For JND-based luminance masking, the parabolic piecewise function, denoted as $L(\mu_Y)$, which also constitutes the luma JND visibility threshold, is employed as a weight to perceptually increase the luma QStep. Function $L(\mu_Y)$ is computed in (1):

$$L(\mu_Y) = \begin{cases} a \cdot \left(1 - \dfrac{2\mu_Y}{2^b}\right)^d + 1, & \text{if } \mu_Y \leq \dfrac{2^b}{2} \\ c \cdot \left(\dfrac{2\mu_Y}{2^b} - 1\right)^f + 1, & \text{otherwise} \end{cases} \quad (1)$$

where parameters $a$, $c$, $d$ and $f$ are set to values 2, 0.8, 3 and 2, respectively. In [16], Naccari and Mrak adopt these parameter values based on the research conducted by Zhang et al. in [29]. The empirical findings by Zhang et al. in [29] correspond to the fact that HVS is perceptually less sensitive to compression artifacts within high intensity and low intensity regions in luma data.

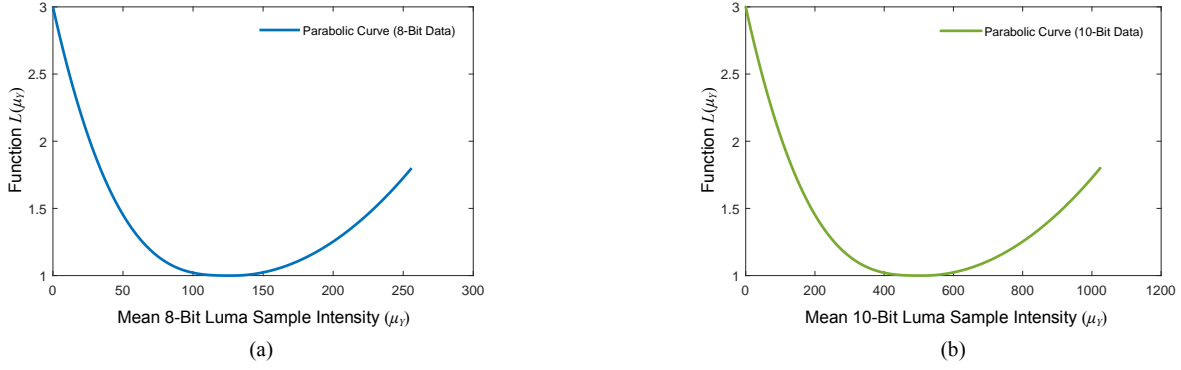

**Fig. 1**: The parabolas derived from function $L(\mu_Y)$ in (1). The subfigures are as follows: (a) corresponds to the parabolic curve when $b = 8$ (8-bit luma data) and (b) corresponds to the parabolic curve when $b = 10$ (10-bit luma data). The integrity of the parabolic curve is preserved irrespective of the bit depth.

Somewhat dissimilar to Eq. (1) in [16], we replace value 256 with $2^b$ in (1) to extend its dynamic range capacity so that it can be used with luma data of any bit depth (see Fig. 1). Variable $\mu_Y$ denotes the mean raw sample in a luma CB; $\mu_Y$ is computed in (2):

$$\mu_Y = \frac{1}{N^2} \sum_{n=1}^{N^2} w_{Y_n} \quad (2)$$

where $N^2$ denotes the number of samples in all CBs (note that the Y CB, the Cb CB and the Cr CB are the same size for 4:4:4 data) and where $w_Y$ refers to the $n^{th}$ raw sample value in a luma CB. Recall that SC-PAQ is designed to modify the URQ-related QSteps at the CB level. The CB-level JND-based perceptual luma QStep and the corresponding perceptual luma QP, denoted as $PStep_Y$ and $PQP_Y$, respectively, are quantified in (3) and (4), respectively:

$$PStep_Y = QStep_Y \cdot \left[ L(\mu_Y) \right] \quad (3)$$

$$PQP_Y = \left[ 6 \times \log_2(PStep_Y) \right] + 4 \quad (4)$$

where $QStep_Y$ denotes the URQ luma QStep in HEVC [12]. For JND-based chrominance masking, the corresponding parabolic piecewise functions for chroma Cb and chroma Cr data are denoted as $C_{Cb}(\mu_{Cb})$ and $C_{Cr}(\mu_{Cr})$, respectively. These functions also act as the Cb and Cr JND visibility thresholds, respectively. $C_{Cb}(\mu_{Cb})$ and $C_{Cr}(\mu_{Cr})$ are computed in (5) and (6), respectively:

$$C_{Cb}(\mu_{Cb}) = \begin{cases} \dfrac{-\mu_{Cb} \cdot (g-1)}{h+g} & \text{if } \mu_{Cb} \leq h \\ 1 & \text{if } h < \mu_{Cb} < j \\ \dfrac{(\mu_{Cb} - j) \cdot (k-1)}{(2^b - 1 - j) + 1}, & \text{otherwise} \end{cases} \quad (5)$$

$$C_{Cr}(\mu_{Cr}) = \begin{cases} \dfrac{-\mu_{Cr} \cdot (g-1)}{h+g} & \text{if } \mu_{Cr} \leq h \\ 1 & \text{if } h < \mu_{Cr} < j \\ \dfrac{(\mu_{Cr} - j) \cdot (k-1)}{(2^b - 1 - j) + 1}, & \text{otherwise} \end{cases} \quad (6)$$

where parameters $g$, $h$, $j$ and $k$ are set to values 3, 85, 90 and 3, respectively. These parameters are adopted in accordance with the chrominance masking-based psychovisual experiments conducted by G. Wang et al. in [22]. Note that the functions in (5) and (6) can be utilized with chroma Cb and Cr data of any bit depth.

In (5) and (6), variables $\mu_{Cb}$ and $\mu_{Cr}$ denote the mean raw sample values in a chroma Cb CB and a chroma Cr CB, respectively. Note that due to the significant similarity of Cb and Cr data in the YCbCr color space, the perceptual masking of chrominance Cb and Cr data can be achieved by virtue of the overlapping effect between chrominance saturation and luminance adaptation [22]. Therefore, high levels of (imperceptible) distortion can be applied to chroma data. Variables $\mu_{Cb}$ and $\mu_{Cr}$ are computed in (7) and (8), respectively:

$$\mu_{Cb} = \frac{1}{N^2} \sum_{m=1}^{N^2} z_{Cb_m} \quad (7)$$

$$\mu_{Cr} = \frac{1}{N^2} \sum_{i=1}^{N^2} s_{Cr_i} \quad (8)$$

where variables $z_{Cb}$ and $s_{Cr}$ refer to the $m^{th}$ raw sample and the $i^{th}$ raw sample in a Cb CB and a Cr CB, respectively. Similar to our work in [33], which is closely related to our work in [34, 35], we exploit the CU-level chroma Cb and Cr CB QP offset signaling mechanism in the Picture Parameter Set (PPS) [36]. This facilitates a straightforward encoder side implementation of SC-PAQ and therefore ensures that the resulting bitstream conforms to the HEVC standard (i.e., ITU-T with Rec. H.265 [3]).

In relation to the initial QPs employed to evaluate SC-PAQ (i.e., QPs 22, 27, 32 and 37), the chroma Cb and Cr QPs are perceptually increased at the CB level by offsetting them against $PQP_Y$. These CB level chroma Cb and Cr QP offsets, denoted as $OQP_{Cb}$ and $OQP_{Cr}$, respectively, are quantified in (9)-(12), respectively.

$$OQP_{Cb} = PQP_Y + PQP_{Cb} \quad (9)$$

$$OQP_{Cr} = PQP_Y + PQP_{Cr} \quad (10)$$

$$PQP_{Cb} = \left[ 6 \times \log_2 \left( QStep_{Cb} \cdot \left[ C_{Cb}(\mu_{Cb}) \right] \right) \right] + 4 \quad (11)$$

$$PQP_{Cr} = \left[ 6 \times \log_2 \left( QStep_{Cr} \cdot \left[ C_{Cr}(\mu_{Cr}) \right] \right) \right] + 4 \quad (12)$$

where $QStep_{Cb}$ and $QStep_{Cr}$ denote the URQ chroma Cb and Cr QSteps, respectively, in HEVC [12] and where [ · ] corresponds to the nearest integer function. As shown in (11) and (12), $C_{Cb}(\mu_{Cb})$ and $C_{Cr}(\mu_{Cr})$ separately adjust $QStep_{Cb}$ and $QStep_{Cr}$, respectively, to arrive at the perceptual chroma QPs. The luma and chroma reconstruction errors, denoted as $q_Y$, $q_{Cb}$ and $q_{Cr}$, respectively, are noticeable only if these values exceed JND visibility thresholds $L(\mu_Y)$, $C_{Cb}(\mu_{Cb})$ and $C_{Cr}(\mu_{Cr})$. Consequently, visually lossless coding is achieved assuming that: $|q_Y| \leq L(\mu_Y)$, $|q_{Cb}| \leq C_{Cb}(\mu_{Cb})$ and $|q_{Cr}| \leq C_{Cr}(\mu_{Cr})$.

**Table 1**: The bitrate reductions (in bold green text) achieved by SC-PAQ compared with anchors: HM 16.10 + SCM 8.0 (left) and IDSQ [16] (right).

Overall Bitrate Reductions (%) Per Sequence and PSNR Value Reductions (dB) Per Channel — RA: Averaged Over QPs 22, 27, 32, 37

| Sequence (YCbCr 4:4:4) | SC-PAQ versus HM 16.10 + SCM 8.0 | | | | SC-PAQ versus IDSQ [16] | | | |
|---|---|---|---|---|---|---|---|---|
| | Bitrate (%) | Y PSNR (dB) | Cb PSNR (dB) | Cr PSNR (dB) | Bitrate (%) | Y PSNR (dB) | Cb PSNR (dB) | Cr PSNR (dB) |
| Basketball Screen | **−13.9%** | −0.45 dB | −2.94 dB | −2.91 dB | **−11.8%** | −0.18 dB | −2.82 dB | −2.81 dB |
| CAD Waveform | **−4.6%** | −1.39 dB | −4.07 dB | −4.58 dB | **−4.4%** | −1.36 dB | −3.82 dB | −4.42 dB |
| Console | **−9.7%** | −0.79 dB | −3.29 dB | −3.68 dB | **−9.1%** | −0.64 dB | −3.08 dB | −3.48 dB |
| Desktop | **−5.5%** | −1.14 dB | −3.80 dB | −3.77 dB | **−5.5%** | −0.90 dB | −3.58 dB | −3.50 dB |
| Flying Graphics | **−12.6%** | −0.44 dB | −3.12 dB | −2.91 dB | **−11.4%** | −0.32 dB | −3.07 dB | −2.89 dB |
| Kimono (10-Bit) | **−48.3%** | −0.13 dB | −0.74 dB | −1.19 dB | **−40.6%** | 0.01 dB | −0.67 dB | −1.08 dB |
| Mission Control 3 | **−11.7%** | −0.55 dB | −2.84 dB | −2.78 dB | **−9.7%** | −0.28 dB | −2.69 dB | −2.62 dB |
| PCB Layout | **−2.1%** | −1.80 dB | −5.30 dB | −6.39 dB | **−2.1%** | −1.33 dB | −5.19 dB | −5.87 dB |
| PPT DOC XLS | **−5.0%** | −0.85 dB | −3.28 dB | −3.29 dB | **−4.3%** | −0.50 dB | −3.00 dB | −3.13 dB |

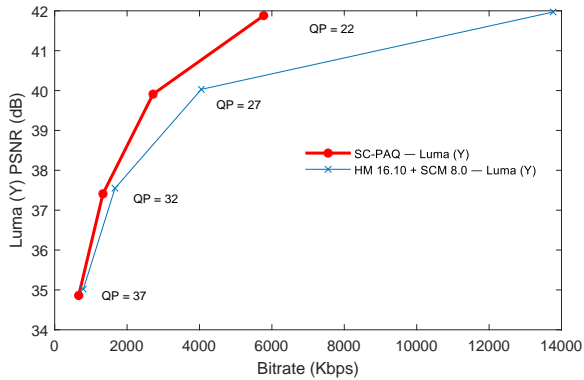

**Fig. 2**: A plot showing the bitrate reductions attained by SC-PAQ (Y channel) over four QP data points compared with HM 16.10 + SCM 8.0 on the Kimono 4:4:4 10-bit sequence (RA encoding configuration).

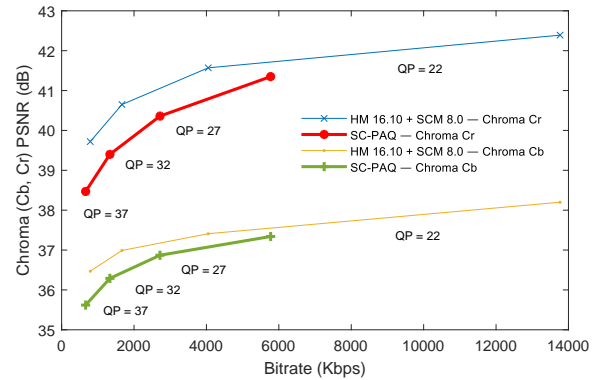

**Fig. 3**: A plot showing the bitrate reductions attained by SC-PAQ (Cb and Cr channels) over four QP data points compared with HM 16.10 + SCM 8.0 on the Kimono 4:4:4 10-bit sequence (RA encoding configuration).

## 4. EVALUATION, RESULTS AND DISCUSSION

SC-PAQ is evaluated and compared with HEVC HM 16.10 + SCM 8.0 (anchor 1) and also Naccari's and Mrak's JND-based IDSQ technique in [16] (anchor 2). SC-PAQ is implemented into HM 16.10 + SCM 8.0 [4] and is subsequently tested on nine official JCT-VC YCbCr 4:4:4 SC sequences; all of the sequences are listed in Table 1. Note that Kimono is the only 10-bit sequence; all other sequences consist of 8-bits per sample.

The experimental setup includes both objective and subjective visual quality evaluations. The objective evaluations correspond to the Common Test Conditions for Screen Content Coding (lossy coding), as recommended by JCT-VC in JCTVC-U1015 [10]. As such, the initial QPs 22, 27, 32 and 37 in addition to the Random Access (RA) encoding configuration and the SCC-Main profile are employed. We follow ITU-R Rec. P.910 in the subjective evaluation procedure. Four individuals engaged in the perceptual assessment; they compared the visual quality of the SC-PAQ coded sequences with the sequences coded by anchor 1, anchor 2 and also the raw video data. The subjective evaluation results are quantified using the Mean Opinion Score (MOS). All of the subjective tests are conducted on a HD 1080p 32 inch display at a close viewing distance of $0.75\text{m} \approx 29.5$ inch; note that this viewing distance is chosen to ensure that all potential compression artifacts are maximally visible.

In terms of the results, the most noteworthy bitrate reductions accomplished by SC-PAQ are attained on the CCC 10-bit sequence Kimono. Bitrate reductions of 48.3% (versus anchor 1) and 40.6% (versus anchor 2) are achieved by SC-PAQ; see Table 1, Fig. 2 and Fig. 3. SC-PAQ attains these vast bitrate reductions due to the high levels of HVS-related psychovisual redundancy in the chroma Cb and Cr CBs within the Kimono 10-bit sequence.

In the subjective evaluations, a total of 106 visual inspections were conducted by the participants. In all of the tests in which the sequences coded by SC-PAQ were compared with the sequences coded by anchors (see Table 1), the participants confirmed that no perceptual visual quality differences could be discerned. More specifically, an MOS = 5 was chosen by the participants in all tests; note that MOS = 5 means 'imperceptible'.

Compared with the visual quality of all uncompressed raw video sequences, the subjective evaluation participants confirmed that the SC-PAQ coded sequences (in all of the RA QP = 22 tests) are perceptually identical to the raw video data. More specifically, in spite of the considerably inferior mathematical reconstruction quality incurred by SC-PAQ, as quantified by reductions in Y, Cb and Cr PSNR dB values, an MOS = 5 (i.e., visually lossless coding) was chosen by each participant in all of the RA QP = 22 tests. In terms of the visually lossless coding test on the Kimono 10-bit sequence (RA QP = 22 test), the SC-PAQ bitstream file size is 2,938 Kilobytes (KB), whereas the raw video data is 2,916,000 KB in size. Furthermore, the mathematically lossless coded version (coded by anchor 1) is 270,713 KB in size.

## 5. CONCLUSION

We propose a JND-based perceptual quantization technique for HM RExt + SCM named SC-PAQ. Both luminance masking and chrominance masking are exploited in order to achieve a high compression performance. In comparison with anchor 1 and anchor 2, SC-PAQ achieves bitrate reductions of up to 48.3% and 40.6%, respectively. In addition, when applied to high bit depth YCbCr 4:4:4 SC data, SC-PAQ achieves visually lossless coding at very low bitrates. Moreover, SC-PAQ attains marginal encoding time and decoding time reductions compared with anchors.

## REFERENCES

[1] D. Flynn, D. Marpe, M. Naccari, T. Nguyen, C. Rosewarne, K. Sharman, J. Sole and J. Xu, "Overview of the Range Extensions for the HEVC Standard: Tools, Profiles, and Performance," *IEEE Trans. Circuits Syst. Video Techn.*, vol. 26, no. 1, pp. 4-19, 2016.

[2] J. Xu, R. Joshi, and R. A. Cohen, "Overview of the Emerging HEVC Screen Content Coding Extension," *IEEE Trans. Circuits Syst. Video Techn.*, vol. 26, no. 1, pp. 50-62, 2016.

[3] ITU-T/ISO/IEC, "ITU-T Rec. H.265/HEVC (Version 4) | ISO/IEC 23008-2, Information Technology – Coding of Audio Visual Objects," 2016.

[4] Joint Collaborative Team on Video Coding. JCT-VC HEVC HM RExt SCM Reference Software, HEVC HM 16.10 + SCM 8.0. Available: http://hevc.hhi.fraunhofer.de/

[5] X. Xu, S. Liu, T. Chuang, Y. Huang, S. Lei, K. Rapaka, C. Pang, V. Seregin, Y. Wang, and M. Karczewicz, "Intra Block Copy in HEVC Screen Content Coding Extensions," *IEEE J. Emerg. Sel. Topics Circuits Syst.*, vol. 6, no. 4, pp. 409-419, 2016.

[6] W. Pu, M. Karczewicz, R. Joshi, V. Seregin, F. Zou, J. Solé, Y. Sun, T. Chuang, P. Lai, S. Liu, S. Hsiang, J. Ye, and Y. Huang, "Palette Mode Coding in HEVC Screen Content Coding Extension," *IEEE J. Emerg. Sel. Topics Circuits Syst.*, vol. 6, no. 4, pp. 420-432, 2016.

[7] Z. Wang, J. Zhang, N. Zhang and S. Ma, "Adaptive Motion Vector Resolution Scheme for Enhanced Video Coding," *Data Compression Conf.,* Snowbird, Utah, USA, 2016, pp. 101-110.

[8] L. Zhang, X. Xiu, J. Chen, M. Karczewicz, Y. He, Y. Ye, J. Xu, J. Solé, and W. Kim, "Adaptive Color-Space Transform in HEVC Screen Content Coding," *IEEE J. Emerg. Sel. Topics Circuits Syst.*, vol. 6, no. 4, pp. 446-459, 2016.

[9] L. Zhang, J. Chen, J. Solé, M. Karczewicz, X. Xiu, and J. Xu, "Adaptive Color-Space Transform for HEVC Screen Content Coding," *Data Compression Conf.,* Snowbird, Utah, USA, 2015, pp. 233-242.

[10] H. Yu, R. Cohen, K. Rapaka and J. Xu, "Common Test Conditions for Screen Content Coding," *JCTVC-U1015, 21st Meeting of JCT-VC*, Warsaw, PL, 2015, pp. 1-8.

[11] W. Gao and S. Ma, "Screen Content Coding," in *Advanced Video Coding Systems*, Springer, 2015, pp. 227-230.

[12] G. Sullivan, J-R. Ohm, W. Han and T. Wiegand, "Overview of the High Efficiency Video Coding (HEVC) Standard," *IEEE Trans. Circuits Syst. Video Technol.*, vol. 22, no. 12, pp. 1649-1668, 2012.

[13] M. Karczewicz, Y. Ye and I. Chong, "Rate Distortion Optimized Quantization," *VCEG-AH21 (ITU-T SG16/Q6 VCEG)*, Antalya, Turkey, 2008.

[14] K. McCann, C. Rosewarne, B. Bross, M. Naccari, K. Sharman and G. J. Sullivan (Editors), "HEVC Test Model 16 (HM 16) Encoder Description," *JCTVC-R1002, 18th Meeting of JCT-VC*, Sapporo, JP, 2014, pp. 1-59.

[15] B. Li and J. Xu, "On selective RDOQ," *JCTVC-T0196, 20th Meeting of JCT-VC*, Geneva, CH, 2015, pp. 1-5.

[16] M. Naccari and M. Mrak, "Intensity Dependent Spatial Quantization with Application in HEVC," *IEEE Int. Conf. Multimedia and Expo*, San Jose, USA, 2013, pp. 1-6.

[17] S. Wang, L. Ma, Y. Fang, W. Lin, S. Ma and Wen Gao, "Just Noticeable Difference Estimation for Screen Content Images," *IEEE Trans. Image Processing.*, vol. 25, no. 8, pp. 3838-3850, 2016.

[18] J. Kim, S-H. Bae, and M. Kim, "An HEVC-Compliant Perceptual Video Coding Scheme Based on JND Models for Variable Block-Sized Transform Kernels," *IEEE Trans. Circuits Syst. Video Techn.*, vol. 25, no. 11, pp. 1786-1800, 2015.

[19] S-H. Bae, J. Kim, and M. Kim, "HEVC-Based Perceptually Adaptive Video Coding Using a DCT-Based Local Distortion Detection Probability Model," *IEEE Trans. Image Processing*, vol. 25, no. 7, pp. 3343-3357, 2016.

[20] J. Wu, L. Li, W. Dong, G. Shi, W. Lin and C-C. J. Kuo, "Enhanced Just Noticeable Difference Model for Images with Pattern Complexity," *IEEE Trans. Image Processing*, vol. 26, no. 6, pp. 2682-2693, 2017.

[21] S-H. Bae and M. Kim, "A DCT-Based Total JND Profile for Spatiotemporal and Foveated Masking Effects," *IEEE Trans. Circuits Syst. Video Techn.*, vol. 27, no. 6, pp. 1196-1207, 2017.

[22] G. Wang, Y. Zhang, B. Li, R. Fan and M. Zhou, "A fast and HEVC compatible perceptual video coding scheme using a transform-domain Multichannel JND model," *Multimedia Tools and Applications*, vol. 76, pp. 1-27, 2017.

[23] J. L. Mannos and D. J. Sakrison, "The Effects of a Visual Fidelity Criterion of the Encoding of Images," *IEEE Trans. Information Theory*, vol. 20, no. 4, pp. 525–536, 1974.

[24] A. J. Ahumada and H. A. Peterson, "Luminance Model Based DCT Quantization for Color Image Compression," *Proc. SPIE*, vol. 1666, pp. 365-374, 1992.

[25] A. B. Watson, "DCTune: A Technique for Visual Optimization of DCT Quantization Matrices for Individual Images," *Society for Information Display Digest of Technical Papers*, vol. 24. 1993, pp. 946–949.

[26] C. H. Chou and Y. C. Li, "A Perceptually Tuned Subband Image Coder Based on the Measure of Just Noticeable Distortion Profile," *IEEE Trans. Circuits Syst. Video Technol.*, vol. 5, no. 6, pp. 467-476, 1995.

[27] C. H. Chou and C. W. Chen, "A Perceptually Optimized 3-D Subband Codec for Video Communication Over Wireless Channels," *IEEE Trans. Circuits Syst. Video Techn.*, vol. 6, no. 2, pp. 143-156, 1996.

[28] X. Yang, W. S. Ling, Z. Lu, E. P. Ong, and S. Yao, "Just Noticeable Distortion Model and its Applications in Video Coding," *Signal Processing: Image Communication*, vol. 20, no. 7, pp. 662-680, 2005.

[29] X. Zhang, W. Lin, and P. Xue, "Improved Estimation for Just-Noticeable Visual Distortion," *Signal Processing*, vol. 85, no. 4, pp. 795-808, 2005.

[30] Y. Jia, W. Lin, and A. A. Kassim, "Estimating Just Noticeable Distortion for Video," *IEEE Trans. Circuits Syst. Video Technol.*, vol. 16, no. 7, pp. 820-829, 2006.

[31] Z. Wei and K. N. Ngan, "Spatio-Temporal Just Noticeable Distortion Profile for Grey Scale Image/Video in DCT Domain," *IEEE Trans. Circuits Syst. Video Technol.*, vol. 19, no. 3, pp. 337–346, 2009.

[32] Z. Chen and C. Guillemot, "Perceptually-Friendly H.264/AVC Video Coding Based on Foveated Just-Noticeable-Distortion Model," *IEEE Trans. Circuits Syst. Video Techn.*, vol. 20, no. 6, pp. 806-819, 2010.

[33] L. Prangnell, M. Hernández-Cabronero and V. Sanchez, "Coding Block-Level Perceptual Video Coding for 4:4:4 Data in HEVC," *IEEE Int. Conf. Image Processing* (In Press), Beijing, China, 2017.

[34] L. Prangnell, M. Hernández-Cabronero and V. Sanchez, "Cross-Color Channel Perceptually Adaptive Quantization for HEVC," *Data Compression Conf.* (In Press), Snowbird, Utah, USA, 2017.

[35] L. Prangnell, "Visually Lossless Coding in HEVC: A High Bit Depth and 4:4:4 Capable JND-Based Perceptual Quantisation Technique for HEVC," *Elsevier Signal Processing: Image Communication* (In Press), 2018.

[36] D. Flynn, N. Nguyen, D. He, A. Tourapis, G. Cote and D. Singer, "RExt: CU Adaptive Chroma QP Offsets," in *JCTVC-O0044, 15th Meeting of JCT-VC*, Geneva, CH, 2013, pp. 1-4.